\documentclass[aps,prl,twocolumn,showpacs,superscriptaddress]{revtex4}
\pdfoutput=1 \usepackage{graphicx}

\begin{document}

\title{Signatures of the superfluid-insulator phase transition in
laser driven dissipative nonlinear cavity arrays}

\author{A. Tomadin} \affiliation{NEST-CNR-INFM and Scuola Normale
Superiore, Piazza dei Cavalieri 7, I-56126 Pisa, Italy}

\author{V. Giovannetti} \affiliation{NEST-CNR-INFM and Scuola Normale
Superiore, Piazza dei Cavalieri 7, I-56126 Pisa, Italy}

\author{R. Fazio} \affiliation{NEST-CNR-INFM and Scuola Normale
Superiore, Piazza dei Cavalieri 7, I-56126 Pisa, Italy}

\author{D. Gerace} \affiliation{CNISM and Dipartimento di Fisica
``A. Volta,'' Universit\`a di Pavia, 27100 Pavia, Italy}

\author{I. Carusotto} \affiliation{Institute for Quantum Electronics,
ETH Z\"urich, 8093 Zurich, Switzerland} \affiliation{BEC-CNR-INFM and
Dipartimento di Fisica, Universit\`a di Trento, I-38050 Povo, Italy}

\author{H.E. T\"ureci} \affiliation{Institute for Quantum Electronics,
ETH Z\"urich, 8093 Zurich, Switzerland}

\author{A. Imamoglu} \affiliation{Institute for Quantum Electronics,
ETH Z\"urich, 8093 Zurich, Switzerland}

\date{\today}

\begin{abstract} We analyze the non-equilibrium dynamics of a gas of
interacting photons in an array of coupled dissipative nonlinear
cavities when driven by a pulsed external coherent field.  Using a
mean-field approach, we show that the response of the system is
strongly sensitive to the underlying (equilibrium) quantum phase
transition from a Mott insulator to a superfluid state at commensurate
filling.  We find that the coherence of the cavity emission after a
quantum quench can be used to determine the phase diagram of an
optical many-body system even in the presence of dissipation.
\end{abstract}


\maketitle

\begin{figure}
\includegraphics[width=0.8\linewidth]{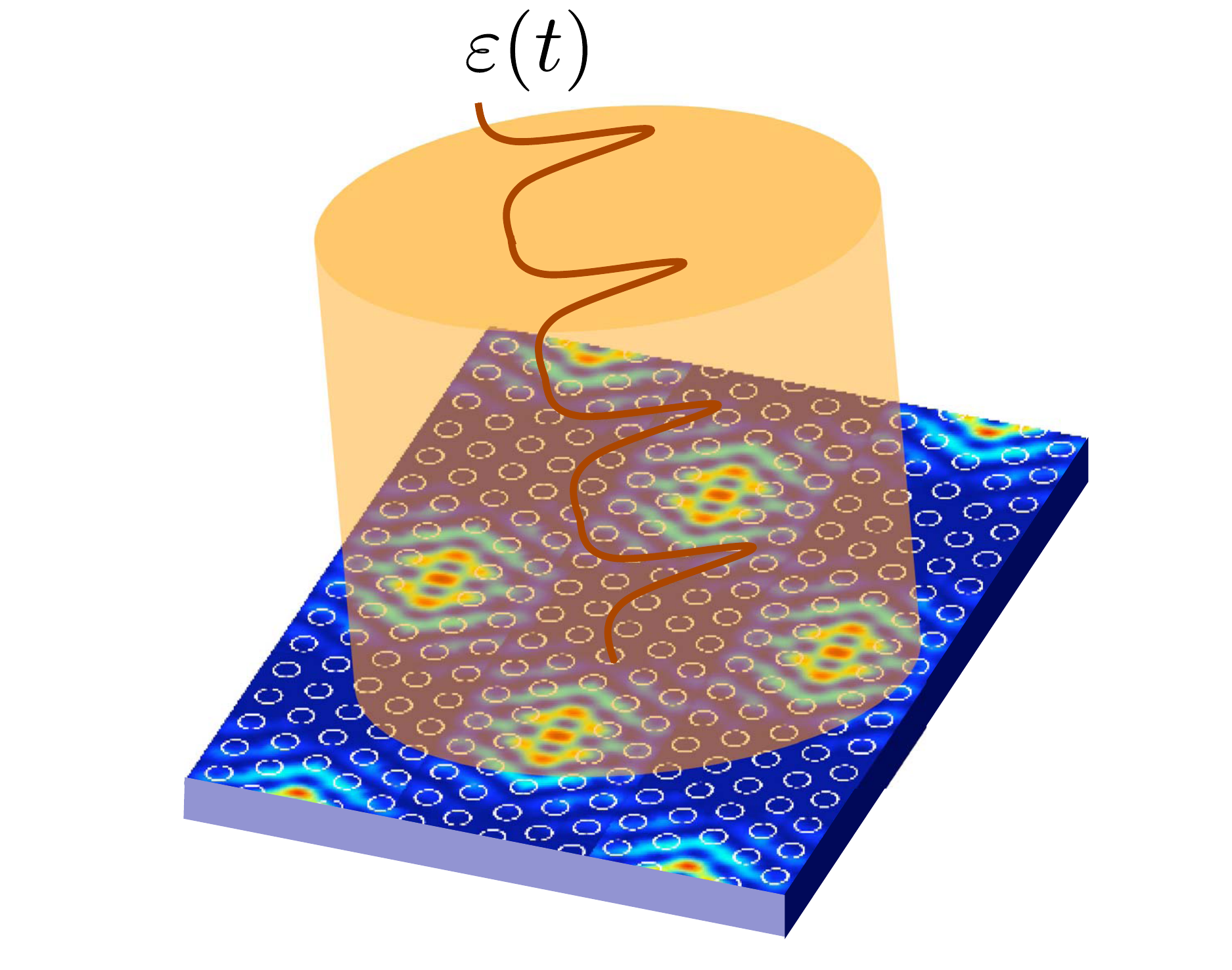}
\caption{\label{fig:array} (Color online) Schematic representation of
  a possible realization of the proposed system, employing a
  2-dimensional array of photonic crystal defect
  cavities~\cite{altug}.  Photons in the cavities have a finite
  lifetime, which requires losses to be compensated by some external
  pump, in our case a coherent pulsed laser field.  Nonlinear
  processes in each cavity lead to an effective Bose-Hubbard repulsion
  between the cavity photons~\cite{hartmann08}.  Neighboring cavities
  are tunnel coupled to each other by spatial overlap of the cavity
  mode profiles.}
\end{figure}

Recent theoretical advances in cavity Quantum Electrodynamics (QED)
have indicated arrays of coupled nonlinear cavities as potential
candidates to explore quantum many-body phenomena of
light~\cite{hartmann08}.  Initial proposals for realizing a Mott phase
of polaritons~\cite{QEDarrays} have been scrutinized in great
detail~\cite{diagram} and different schemes have been put forward to
simulate a variety of correlated quantum
states~\cite{hartmann07,kay08,cho08,carusotto08}.  Given the
importance of dissipation in state-of-the-art solid-state (single)
cavity QED devices (see for example \cite{kevin07nat}), experiments in
cavity QED arrays are expected to be performed under strongly
non-equilibrium conditions, with an external source compensating for
the loss of photons (see Fig.~\ref{fig:array}).  These conditions,
dictated by the experimental constraints, put forward QED arrays as
natural candidates to explore the rich world of non-equilibrium
quantum many-body systems \cite{chang07,josephson,carusotto04}, that
is the subject of recent interest in the context of ultracold atoms as
well~\cite{diehl08,daley09}.  At the same time new important questions
arise related to whether or not it is possible to realize and detect,
under realistic non-equilibrium conditions, the very rich phase
diagram \cite{hartmann08} predicted for QED arrays at equilibrium.

In this Letter we propose an answer to this last question thus
providing the missing link between the initial ideas of cavity arrays
as quantum simulators and future experimental realizations.
Specifically we will show how to detect the superfluid-insulator phase
transition for a dissipative nonlinear cavity array driven by an
external pulsed laser.  In analogy to what happens in a quantum
quench~\cite{AA2002,non-eq}, photons are excited in the cavities by a
periodic train of short, coherent pulses, and then evolve according to
the complex many-body dynamics determined by the simultaneous action
of hopping, strong interactions and losses.  The properties of the
secondary emission are measured in the time lapse between subsequent
pulses: even though the losses drive the system towards the vacuum
state, distinct signatures of the quantum phase transition are found
in the statistics of the transient emitted light.  First- and
second-order photon correlation are shown to be powerful tools to
detect the existence of a non-zero order parameter as they are very
sensitive to the delocalization of photons through the array, although
more complete measurement schemes exist \cite{measure}.

The time evolution of the array in the presence of driving and losses
is described by the master equation
\begin{equation}\label{eq:liouville}
  \partial_{t}\rho(t) = -i[\hat{\cal H},\rho] + \cal{L}[\rho]~,
\end{equation}
for the density matrix $\rho$ of the system.  The first term on the
r.h.s.\ of Eq.~(\ref{eq:liouville}) accounts for the unitary evolution
of the system while the second term accounts for the damping.  As
discussed in~\cite{hartmann08}, an array of coupled cavities in the
absence of losses can realize an effective Bose-Hubbard model whose
Hamiltonian reads
\begin{eqnarray}\label{eq:bosehubbard} \hat{\cal H} = &-&J
  \sum_{\langle \ell \ell' \rangle} \hat{a}_{\ell}^{\dag}
  \hat{a}_{\ell'} + \frac{U}{2} \sum_{\ell} \hat{n}_{\ell}
  (\hat{n}_{\ell} - 1) \nonumber \\ &+&\Delta \sum_{\ell} \hat{n}_{\ell}
  + \sum_{\ell} [\varepsilon (t)\,\hat{a}_{\ell} + \varepsilon
  (t)^{\ast}\,\hat{a}_{\ell}^{\dag}]~,
\end{eqnarray}
with $\hat{a}_{\ell}$, $\hat{a}_{\ell}^{\dagger}$, and
$\hat{n}_{\ell}$ denoting, respectively, the annihilation, creation
and number operator of the electromagnetic mode of the $\ell$th
cavity.  The first term of Eq.~(\ref{eq:bosehubbard}), quantified by
the coupling constant $J$, describes the hopping of photons between
neighboring cavities (the lattice has a coordination number $z$).  The
second term, quantified by $U$, represents the effective nonlinearity
arising from the specific light-matter interaction mechanism.  The
achievable values of $J$ and $U$ depend on the specific implementation
of the model~\cite{hartmann08}.  The last two terms in
Eq.~(\ref{eq:bosehubbard}) describe the coupling of the photons in the
cavities with a uniform coherent pump in the rotating frame -- here
$\varepsilon (t)$ is the (slowly-varying) envelope of the external
driving field and $\Delta$ is the detuning of the laser frequency from
the bare cavity resonance.  The dominant source of dissipation in the
system is the leakage of photons out of the cavities.  In the Markov
approximation, the decay process is described by a Liouvillian in the
Lindblad form ${\cal L}[\rho] = \kappa \sum_{\ell}
(2\hat{a}_{\ell}\rho \hat{a}_{\ell}^{\dag} - \hat{n}_{\ell} \rho -
\rho \hat{n}_{\ell})$, where $(2\kappa)^{-1}$ is the photon lifetime.

Solving Eq.~(\ref{eq:liouville}) is a formidable task, as it describes
an open quantum many-body system out of equilibrium.  We thus resort
to a self-consistent cluster mean-field approach, which should give
reliable results if the array is (at least) two-dimensional.  The
cluster mean-field (for its equilibrium version see for
example~\cite{cluster}) is based on approximating the
Hamiltonian~(\ref{eq:bosehubbard}) with that of a cluster of cavities
that interact with the rest of the lattice via a mean-field term,
i.e. $ \hat{\cal H}_{\rm MF} = \sum_{\ell,\ell' \in \Omega_{\rm c}}
\hat{\cal H}_{\ell,\ell'} - J\sum_{\langle \ell',\ell \rangle, \ell
\in \Omega_{\rm b}, \ell' \notin \Omega_{\rm c}}
[\hat{a}_{\ell}^{\dag} \psi_{\ell'} + \mbox{H.c.}]$. Here, $\hat{\cal
H}_{\ell,\ell'}$ is the Hamiltonian (\ref{eq:bosehubbard}) restricted
to the sites within the cluster ($\Omega_{\rm c}$).  The second term
instead is the mean-field expression for the hopping from the cavities
$\ell$ in the cluster boundary ($\Omega_{\rm b}$) to their
nearest-neighbors $\ell'$ outside the cluster.  It is a function of
the time-dependent mean-field order parameter $\psi_{\ell}(t)$ which,
for the $\ell$th cavity, is determined by the self-consistency
condition $\psi_{\ell}(t) = {\rm Tr} [\hat{a}_{\ell}\,\rho(t)]$.  In
the remainder of the paper we will consider only the two cases in
which the cluster consists of one or two sites.  For the numerical
integration of the master equation we use a fourth order Runge-Kutta
algorithm and truncate the local Fock basis $\{| n \rangle_{\ell}\}_{n
= 0}^{n_{\rm max}}$ for the $\ell$th cavity, with an upper cutoff
$n_{\rm max} \simeq 20 $.

In order to simplify the presentation, we first consider an idealized
case in which the laser pulse sets, at $t=0$, the cavity array in a
Fock state $\rho_{\rm in} = \bigotimes_{\ell} |1 \rangle_{\ell}
\langle 1|_{\ell}$ with one photon per cavity.  We will discuss the
precise shape of the required pulses and the robustness of the
response of the system to imperfections in the pulse shape further
below.  We first focus on single-site mean field and analyze the
coherence properties of the light emitted from the cavity array after
each pulse.  The time-dependence of the photon population $n(t)$ of
each cavity is trivial, the corresponding equation of motion can be
integrated yielding an exponential decay $n(t)= n(0) e^{-2 \kappa t}
$.  Much more interesting are the properties of the order parameter
$\psi_{\ell}(t)$ and the zero-time delay second-order correlation
function, $g_{2} = \langle \hat{n}^{2}_{\ell} - \hat{n} _{\ell}\rangle
/ \langle \hat{n}_{\ell} \rangle^{2}$, for $t > 0$, i.e. after the
laser pulse is switched off.  If $J/U < (J/U)_{\rm c}$ ($(J/U)_{\rm
  c}$ is the value at which the transition from the superfluid to the
Mott state occurs) any initial fluctuation in the order parameter does
not grow in time and the density matrix in photon-number
representation is essentially diagonal throughout the relaxation
process.  In the opposite case, $J/U > (J/U)_{\rm c}$, a superfluid
instability can develop and leads to nonlinear oscillations of both
$|\psi(t)|$ and $g_{2}$.  This kind of instability has been discussed,
in the absence of dissipation, in~\cite{AA2002}. Here we show that the
instability is present also in the presence of losses.  It can be
generated by means of pulsed laser and detected by measuring the
properties of the light emitted from the cavities.  On the contrary,
within our mean-field analysis, continuous shining of coherent light
on the cavities would wash out the effects of the instability.

\begin{figure}
\includegraphics[width=\linewidth]{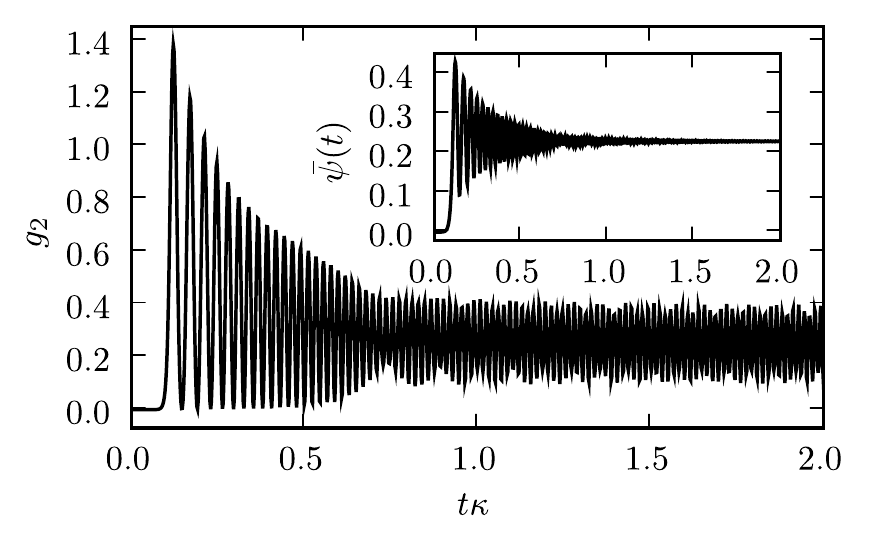}
\caption{\label{fig:one} Time evolution of the zero-time delay
  second-order correlation function and of the rescaled order
  parameter (inset), after a pulse that projects the cavities into a
  Fock state with one photon per cavity.  The rescaled order parameter
  $\bar{\psi}(t)=|\psi(t)| / \sqrt{n(t)}$ compensates for the decay of
  the photon number in the cavities and is related to the coherent
  fraction of the emission ($\kappa=10^{-2}\,U$, $zJ=3.0\,U$,
  $\Delta$=0). }
\end{figure}

For the open system we are considering, a generic example of the
instability induced by the pulsed pump is shown in Fig.~\ref{fig:one},
where we show the evolution of $g_2(t)$ and (normalized) $|\psi(t)|$
calculated for $t > 0$.  The dynamics is characterized by three
different regimes.  At short times, a linear instability sets in and
both quantities increase exponentially, notwithstanding the slow decay
of the photon population.  At intermediate times, the collective
nonlinear dynamics of the array leads to oscillations.  In the long
time limit, these oscillations damp out with a time constant of the
order of $\kappa^{-1}$ (in this regime, when $n(t) \ll 1$, it can be
shown that $|\psi(t)| \sim \sqrt{n(t)}$).

An experimentally measurable observable that yields a clear signature
of the different regimes in the initial transient is the zero-time
delay second-order correlation function averaged over a certain
interval of time.  To calculate such time averaged observables, we
integrate the equation of motion Eq.~(\ref{eq:liouville}) up to the
time $2.0\,\kappa^{-1}$, and determine the time averaged values
$\langle |\psi(t)| \rangle_t$ and $\langle g_{2} \rangle_t$ in the
time interval $1.0\,\kappa^{-1} < t < 2.0\,\kappa^{-1}$.  Considering
time-integrated quantities allows to relax the experimental
requirements on the time resolution of the measurements, which can
hardly exceed $\kappa^{-1}$ in realistic state-of-the-art devices.

\begin{figure}
\includegraphics[width=\linewidth]{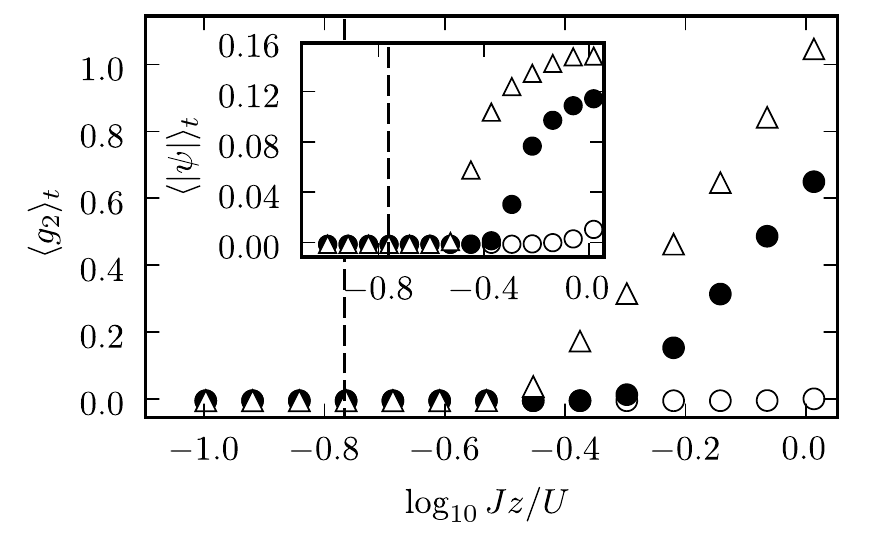}
\caption{\label{fig:two} Time-average of $g_{2}(t)$ (main panel) and
  integral of $|\psi(t)|$ (inset), in the time interval
  $1.0\,\kappa^{-1} < t < 2.0\,\kappa^{-1} $ for different values of
  the damping: $\kappa=2.0\times 10^{-2}\,U$ (empty circles),
  $1.0\times 10^{-2}\,U$ (filled circles), and $0.5\times 10^{-2}\,U$
  (empty triangles); $\Delta=0$.  The vertical dashed line marks the
  critical value of the Mott-to-superfluid transition for the
  equilibrium Bose-Hubbard model ($(zJ/U)_c \sim 0.17$). }
\end{figure}

In Fig.~\ref{fig:two} we show $\langle |\psi(t)| \rangle_t$ and
$\langle g_{2} \rangle_t$ as a function of the ratio $J/U$, for
different values of the dissipation rate.  Both quantities vanish if
$J/U$ is smaller than the critical value, meaning that the order
parameter does not develop any instability, and consequently the light
emitted from the cavities is strongly antibunched.  In the opposite
case both $\langle |\psi(t)| \rangle_t$ and $\langle g_{2} \rangle_t$
are different from zero.  The time-averaged $g_{2}$ as a function of
$J/U$ monotonically rises from zero to almost unity, thus showing a
crossover from antibunched to Poissonian statistics.  Integrating the
equation of motion for times larger than $2.0\,\kappa^{-1}$ leaves the
features depicted in Fig.~\ref{fig:two} qualitatively unaltered.  We
emphasize that the crossover of both $\langle |\psi(t)| \rangle_t$ and
$\langle g_{2} \rangle_t$ takes place in a very narrow range of $J/U$
values around the transition point of the closed system.  The
sensitivity of the light statistics to the coupling between
neighboring cavities is a signature of the many-body origin of this
phenomenon.  The second-order correlation is thus an excellent
candidate to detect the different quantum phases in cavity arrays.

A linear stability analysis, valid for $t\,U,t\,\kappa \ll 1$, shows
that the rate of the initial exponential build-up of the order
parameter is reduced by a quantity $\kappa$, at fixed $J/U$; for large
enough $\kappa$, the superfluid instability is then entirely
suppressed.  As a consequence, the critical ratio $(J/U)_{\rm c}$
increases by $\simeq \kappa^{2} / (2 U^{2})$.  Such trend is already
visible for $U / \kappa \simeq 10^{2}$.  We note that this suppression
of the superfluid phase is not a consequence of the coupling to a bath
\cite{diehl08}.  It is important to note that $U / \kappa \simeq
10^{2}$ is experimentally achievable in a solid-state cavity with a
linewidth of $\simeq 10$ $\mu$eV (corresponding to $Q\simeq 10^5$ at
optical or near-infrared frequencies) and interaction strength $\sim
1$ meV, which are within reach in current solid-state cavity QED
systems~\cite{CombrieVerger}.  In the case of coupled photonic crystal
cavities (see Fig.~\ref{fig:array}) the range of parameters that one
can expect (assuming $U = 0.1$~meV) is $0 < J/U < 50$~\cite{values}.
Alternatively, circuit-QED also provides an ideal experimental
realization for the present proposal~\cite{schuster07nat}.

All the results presented here are robust even when some of the
assumptions made so far are relaxed.  In the vicinity of the critical
point, the order parameter evolves on time scales which are much
longer than the cavity parameters~\cite{AA2002}.  In contrast, the
correlation function might show a more complex short-distance
evolution which could invalidate the analysis presented above.  In
order to check this possibility, we carried out the same analysis in
the two-site mean field approximation, where both the dynamics of the
order parameter and the short time scale dynamics governed by $J$ and
$U$ are present.  All the results shown so far are fully confirmed.
In addition, we checked that the uniform solution for the order
parameter is stable against possible spatial inhomogeneities of the
initial state.

\begin{figure}
\includegraphics[width=\linewidth]{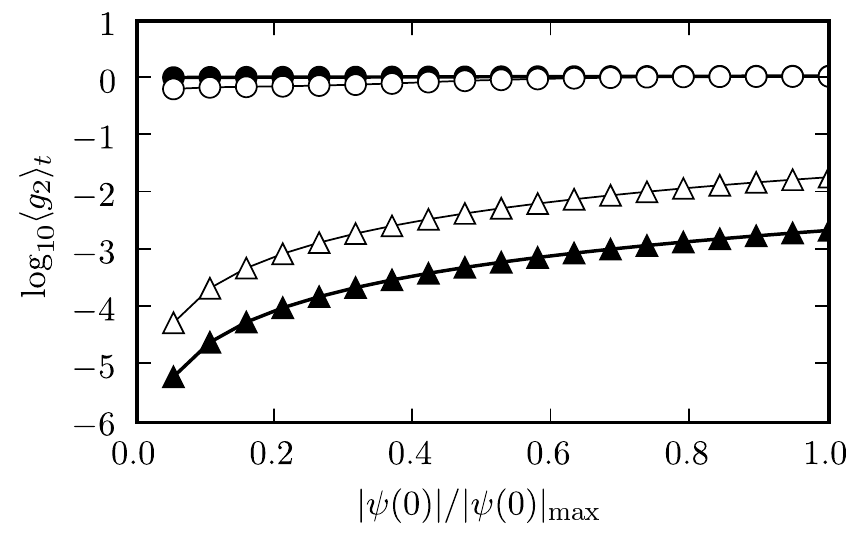}
\caption{\label{fig:three} Time average of $g_{2}(t)$, for $\kappa =
  10^{-2}\,U$.  The population $\rho_{0}$ of the vacuum in the initial
  state is $\rho_{0}=0.02$ for the filled symbols and $\rho_{0}=0.2$
  for the empty symbols.  ($zJ/U=0.1$ (triangles) and $zJ/U=1.0$
  (circles), below and above the critical value; $\Delta=0$.) }
\end{figure}

We also checked that our findings do not depend on the exact form of
the state after the pulsed excitation.  We considered an initial state
that is not necessarily pure, but that entirely projects the system
onto the subspace spanned only by the two Fock states \ $|0
\rangle_{\ell}$ and $|1 \rangle_{\ell}$.  We considered depletion of
the average filling up to $20\%$ and scanned the whole range of
initial coherences.  For all the possible initial states, $\langle
g_{2} \rangle_t$ in the superfluid regime was always markedly
different (at least three orders of magnitude) from the insulating
one, as shown in Fig.~\ref{fig:three}.  This insensitivity to the
initial conditions in the superfluid regime is due to the nonlinearity
of the time-evolution for times shorter than the photon lifetime.  In
fact, in the insulating regime the initial correlations are suppressed
due to the photon blockade, while in the superfluid region even the
absence of any initial correlation is quickly compensated by the
cooperative action of the photons.  We remark that while $\langle
g_{2} \rangle_t$ predicts strongly antibunched cavity emission in the
insulating regime for all considered variations in the initial states,
the actual degree of antibunching does depend on the initial state
coherence; at present, we do not understand this feature.

\begin{figure}
\includegraphics[width=\linewidth]{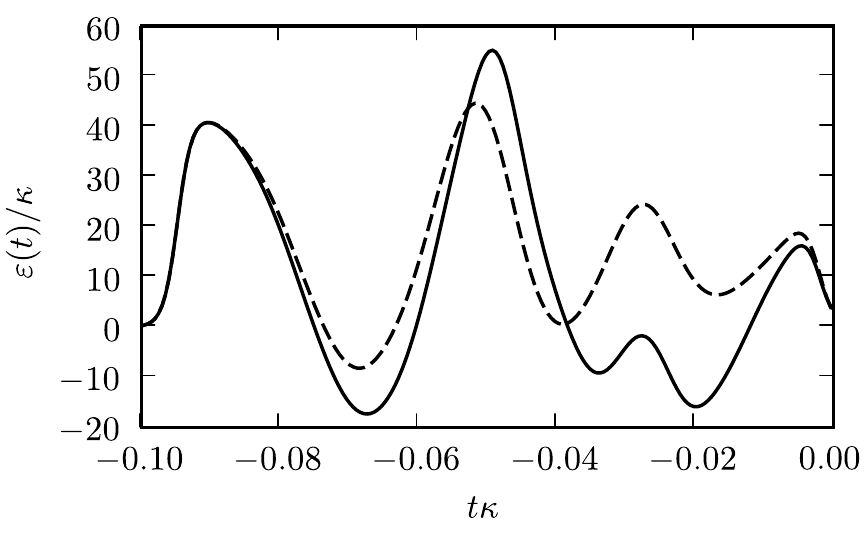}
\caption{\label{fig:four} Optimized pulse envelopes for the
  preparation of a Fock state with one photon per cavity, with a
  fidelity ${\cal F}\simeq 99\%$.  The dashed line is the pulse
  $\varepsilon^{(0)}(t)$ for $J=0$.  The solid line is the pulse
  $\varepsilon^{(1)}(t)$, detuned by $\Delta=43.25\,\kappa $, with
  $zJ/U=0.5$.  The duration $T_{\rm p}=0.1/\kappa $ of the pulse is
  much shorter than the photon lifetime ($\kappa = 10^{-2}\,U$ in the
  plot). }
\end{figure}

We now discuss pulse shapes that project the coupled cavity array into
a desired initial state (such as $\rho_{\rm in}$ discussed before).  A
$\pi$-pulse applied to an isolated cavity, drives the vacuum state
into the desired initial state, provided that the duration of the
pulse is shorter than the photon lifetime $\kappa^{-1}$.  To design a
suitable pulse shape for the whole array, we resorted to a
rapidly-converging quantum optimal control algorithm \cite{WG2007}:
first, we found an optimized envelope $\varepsilon^{(0)}(t)$ (see
Fig.~\ref{fig:four}, dashed line) for a single cavity ($J=0$).  As
dissipation does not play a substantial role during the application of
the pulse in the time interval $-T_{\rm p} < t < 0$, we use an
algorithm of quantum optimal control for pure states, obtaining a
fidelity ${\cal F}\simeq 99\%$ with respect of the desired state
$\rho_{\rm in}$.  In the case of an array of coupled cavities, $J \ge
0$ substantially modifies the response of the system to the pumping
field and it is necessary to devise a different shape
$\varepsilon^{(1)}(t)$ of the pulse for each value of the hopping
amplitude (see Fig.~\ref{fig:four}, solid line).  It is nevertheless
possible to prepare the initial state with high fidelity ($99\%$ in
Fig.~\ref{fig:four}) in a substantial range of values up to $zJ
\lesssim U$.

In conclusion, we have shown that the coherence properties of the
secondary cavity emission, induced by a pulsed excitation, provide
signatures of the collective many-body phase of the array of nonlinear
cavities.  Our analysis fully takes into account the intrinsically
driven-dissipative nature of the strongly correlated quantum many-body
system and identifies how dissipation influences the underlying
equilibrium quantum phase transition that this system exhibits in the
absence of losses.  Coupled cavities are promising candidates to
simulate open quantum many-body systems and a study of their
non-equilibrium behavior will certainly unveil a variety of new
phenomena.

\begin{acknowledgements} We acknowledge useful discussions with
Sebastian Schmidt. This work was supported by EU IP-EuroSQIP (RF), the
Swiss NSF under Grant No. PP00P2-123519/1 (HET) and an ERC Advanced
Investigator Grant (AI).
\end{acknowledgements}

\end{document}